\input harvmac
\global\newcount\fno \global\fno=0
\def\myfoot#1{\global\advance\fno by1
\footnote{$^{\the\fno}$}{#1}}

\def\pl#1{{\it Phys. Lett.} {\bf #1B}}
\def\prl#1{{\it Phys. Rev. Lett.} {\bf #1}}
\def\prd#1{{\it Phys. Rev.} {\bf D#1}}

\def\np#1{{\it Nucl. Phys.} {\bf B#1}}

\def\jmp#1{{\it J. Math. Phys.} {\bf #1}}

\def\p{{\scriptscriptstyle +}}
\def\m{{\scriptscriptstyle -}}


\def\ha{
\hbox to\msize{\hfil${{\scriptstyle 1}\over{\scriptstyle 2}}$}}
\def\mha{
\hbox to\msize{\hfil--${{\scriptstyle 1}\over{\scriptstyle 2}}$}}
\def\qt{
\hbox to\msize{\hfil${{\scriptstyle 1}\over{\scriptstyle 4}}$}}
\def\mqt{
\hbox to\msize{\hfil--${{\scriptstyle 1}\over{\scriptstyle 4}}$}}
\def\mtq{
\hbox to\msize{\hfil--${{\scriptstyle 3}\over{\scriptstyle 4}}$}}
\def\ze{\hbox to\msize{\hfil$0$}}
\def\one{\hbox to\msize{\hfil$1$}}
\def\mone{\hbox to\msize{\hfil--$1$}}

\def\roughly#1{\raise.3ex\hbox{$#1$\kern-.75em\lower1ex\hbox{$\sim$}}}

\def\p{\raise2pt\hbox to5pt{\fiverm +}}
\def\m{\raise2pt\hbox to5pt{\fiverm --}}

\def\pl#1{{\it Phys. Lett.} {\bf #1B}}
\def\prl#1{{\it Phys. Rev. Lett.} {\bf #1}}
\def\prd#1{{\it Phys. Rev.} {\bf D#1}}

\def\np#1{{\it Nucl. Phys.} {\bf B#1}}

\def\jmp#1{{\it J. Math. Phys.} {\bf #1}}

\def\darr#1{\raise1.5ex\hbox{$\leftrightarrow$}\mkern-16.5mu #1}

\catcode`@=11
\def\myeqalign#1{\null\,\vcenter{\openup-1\jot \m@th
 \ialign{\strut$\displaystyle{##}$&$\displaystyle{{}##}$\hfil
	\crcr#1\crcr}}\,}
\def\roughly#1{\raise.3ex\hbox{$#1$\kern-.75em\lower1ex\hbox{$\sim$}}}
\def\ap{\alpha ^{\prime}}
\def\mp{M_{\rm P}}
\def\ag{\alpha _{\rm U}}
\def\l{\lambda_{I}}
\def\op{${\rm I}^{\prime}$}
\def\lh{\lambda_h}
\def\lp{\lambda_{I^{\prime}}}
\def\va{v_{\alpha}}
\def\vat{{\tilde v}_{\alpha}}
\def\etp{{\rm e}^{-2\phi}}
\def\Tr{{\rm Tr}\,}
\def\tr{{\rm tr}\,}
\def\tri{{\rm tr}_i\,}
\def\trj{{\rm tr}_j\,}
\overfullrule=0pt
\lref\sch{J. Schwarz,
\pl{360} (1995) 13, erratum \pl{364} (1995) 252;
\pl{367} (1996) 97.}

\lref\witva{P. Horava and E.
Witten, \np{460} (1996) 506.}

\lref\witan{E. Witten, \np{460} (1996) 541.}

\lref\bagking{J. Bagger, C. Schmidt, and S. King,
\prd{37} (1988) 1188.}

\lref\ankk{I. Antoniadis, \pl{246} (1990) 377;
I. Antoniadis, K. Benakli, and M. Quiros,
\pl{331} (1994) 313.}

\lref\farman{A. Faraggi, ``Realistic Superstring Models'',
hep-ph/9405357.}

\lref\keithr{K. Dienes, ``String Theory and the Path
to Unification: a Review of Recent Developments'',
hep-th/9602045.}

\lref\ferr{F. Quevedo, ``Lectures on Superstring
Phenomenology'', hep-th/9603074.}

\lref\mer{J. Lykken, ``Four-Dimensional Superstring
Models'', hep-ph/9511456.}

\lref\misal{K. Dienes, M. Moshe, and R. Myers,
\prl{74} (1995) 4767.}

\lref\witpol{J. Polchinski and E. Witten,
\np{460} (1996) 525.}

\lref\dinshir{M. Dine and Y. Shirman,
``Truly Strong Coupling and Large Radius in String Theory'',
hep-th/9601175.}

\lref\polgim{E. Gimon and J. Polchinski,
``Consistency Conditions for Orientifolds and D-Manifolds'',
hep-th/9601038.}

\lref\fern{G. Aldazabal, A. Font, L. Ibanez, and
F. Quevedo, ``Heterotic/Heterotic Dualit in $D=6$,4'',
hep-th/9602097.}

\lref\green{M. Green, J. Schwarz, and P. West, \np{254} (1985) 327.}

\lref\jorr{J. Lopez, ``Supersymmetry: from the Fermi Scale
to the Planck Scale'', hep-ph/9601208.}

\lref\oldwit{E. Witten, \np{443} (1995) 85.}

\lref\bandin{T. Banks and M. Dine, \prd{50} (1994) 7454.}

\lref\oldsd{M. Dine and N. Seiberg, \pl{162} (1985) 299.}

\lref\swsix{N. Seiberg and E. Witten,
``Comments on String Dynamics in Six Dimensions'',
hep-th/9603003.}

\lref\scharzsix{J. Schwarz, ``Anomaly-free Supersymmetric
Models in Six Dimensions'', hep-th/9512053.}

\lref\polnote{J. Polchinski, S. Chaudhuri,
and C. V. Johnson, ``Notes on D-Branes'',
hep-th/9602052.}

\lref\band{M. Bander, ``Quark Resonances and
high ${\rm E}_{\rm t}$ Jets'', hep-ph/9602330.}

\lref\sagnotti{A. Sagnotti, \pl{294} (1992) 196.}

\lref\cdfjet{CDF collaboration (F. Abe et al),
``Inclusive Jet Cross-Section in $p\bar p$ Collisions
at $\sqrt{s}$$=$$1.8$ TeV'', hep-ex/9601008.}

\lref\bsagnotti{M. Bianchi and A. Sagnotti, \np{361} (1991) 519.}

\lref\erler{J. Erler, \jmp{35} (1994) 1819.}

\lref\dufw{M. J. Duff, R. Minasian and E. Witten,
``Evidence for Heterotic/Heterotic Duality'', hep-th/9601036.}

\lref\ftheo{C. Vafa, ``Evidence for F-Theory'', hep-th/9602022.}

\lref\compf{D. R. Morrison and C. Vafa,
``Compactifications of F-Theory on Calabi-Yau
Threefolds'', hep-th/9602114.}

\lref\newjoe{M. Berkooz, R. Leigh, J. Polchinski, J. Schwarz,
and N. Seiberg, ``Anomalies and Duality in Type I Superstrings
in Six Dimensions'', to appear.}

\lref\newwit{E. Witten, ``Strong Coupling Expansion
of Calabi-Yau Compactification'', hep-th/9602070.}

\lref\mesb{J. Lykken, ``String Models of SUSY GUTs'',
talk presented at the Workshop on SUSY Phenomena and SUSY GUTs,
ITP Santa Barbara, December 7-9, 1995.}

\lref\cchl{ S. Chaudhuri, S.-w. Chung, G. Hockney and J. Lykken,
\np{456} (1995) 89.}

\lref\chl{ S. Chaudhuri, G. Hockney and J. Lykken,
{\it Phys. Rev. Lett.}\/ {\bf 75} (1995) 2274.}

\lref\threeg{ S. Chaudhuri, G. Hockney and J. Lykken,
``Three Generations in the Fermionic Construction'',
hep-th/9510241.}

\lref\gins{P. Ginsparg, \pl{197} (1987) 139.}

\lref\hag{J. Atick and E. Witten, \np{310} (1988) 291.}
\Title{\vbox{\baselineskip12pt\hbox{FERMILAB-PUB-96/070-T}
\hbox{hep-th/9603133}}}
{Weak Scale Superstrings}
\bigskip\vskip2ex
\centerline{{\bf Joseph D. Lykken\footnote{$^*$}
{e-mail : lykken@fnal.gov}}}
\medskip
\centerline{Fermi National Accelerator Laboratory,}
\centerline{P.O. Box 500, Batavia, IL 60510}
\vskip .5cm
\noindent
Recent developments in string duality suggest that
the string scale may not be irrevocably tied to
the Planck scale. Two explicit but unrealistic examples
are described where the ratio of the string scale to the Planck scale is
arbitrarily small. Solutions which are more realistic may exist in the
intermediate coupling or ``truly strong coupling'' region of the
heterotic string. Weak scale superstrings have dramatic experimental
consequences for both collider physics and cosmology.

\Date{3/96}

\newsec{Introduction}

The discovery of string dualities is reshaping the way we
think about string theory. Indeed even the terminology
``string theory'' has become suspect, given the apparent
dualities between certain string compactifications and
compactifications of eleven-dimensional ``$M$-theory''
\refs{\sch,\witva,\witan,\dufw}
or twelve-dimensional ``$F$-theory''
\refs{\ftheo,\compf}. The heterotic, Type II,
Type I, and Type \op\ superstrings are dual
descriptions of the same underlying theory.

In light of these radical developments it is
important to reexamine our understanding of
how string theory is likely to be related to
the real world.
A step in this direction is the recent paper
by Witten \refs{\newwit}. He observes that
superstring phenomenology to date has assumed certain
relationships between parameters which hold in
the weak coupling regime of the heterotic string,
but which may {\it not} be valid generally.
In particular there is the famous tree-level
formula \gins

\eqn\trelev{
\ap \mp ^2 = {4\over k\ag } \quad .
}
Here $\ap$ is the string tension (which has units of
length squared); for simplicity we will define
$m_s = 1$$/$$\sqrt{\ap}$ to be the string scale.
$\mp$ is the Planck mass $\simeq 10^{19}$
GeV defined from Newton's constant by $G_{\rm N} = 1/\mp ^2$.
$\ag = g_{\rm U}^2/4\pi$, where
$g_{\rm U}$ is the unified gauge coupling.
The parameter $k$ is the Kac-Moody level; it
is compactification dependent but of order one
\refs{\mesb}. If the group is nonsimple $k$
takes independent values for each group factor.

Since the value of $g_{\rm U}$ is presumably of order one,
this implies that the string scale $m_s$
is not far below the Planck scale.
The string scale determines both the scale of gauge
coupling unification and the scale of Regge recurrences
(the massive string modes). These are thus both predicted
to be in the range $10^{17}$ -- $10^{18}$ GeV.

Reference \refs{\newwit} points out that this relationship
of scales and couplings can be radically altered
in the strong coupling regime of the heterotic
string. This is shown by a duality map of the strong coupling
$SO(32)$ or $E_8$$\times$$E_8$ heterotic strings,
compactified to four dimensions, to (respectively)
a weak coupling Type I string compactified to four dimensions,
or $M$-theory compactified first to
$R^{10}$$\times$$S^1$$/$$Z_2$, then to four dimensions.
For the $SO(32)$ string one finds that

\eqn\newrela{
m_s^2/\mp ^2 \propto \l
}
where $\l$ is the ten-dimensional Type I string coupling,
determined dynamically by the
vacuum expectation value of the dilaton.
Since we are in the weak coupling regime for
the Type I string \newrela\ can imply
small values of $m_s^2/\mp ^2$.

In the $E_8$$\times$$E_8$ case one finds that

\eqn\newrelb{
m_s^2/\mp ^2 \propto \kappa ^{2/9}/\rho
}
where $\kappa$ is the eleven-dimensional gravitational
coupling and $\rho$ is the compactification radius in
$R^{10}$$\times$$S^1$. Here the story is more complicated,
but in \newwit\ it is shown that,
for the symmetric embedding of the gauge bundle, the
ratio $m_s^2/\mp ^2$ can also be small
consistent with the assumption that the ten-dimensional
fields are weakly coupled.

If the string scale is not irrevocably tied to the
Planck scale, it is natural to explore the idea
that it may instead be tied to the electroweak scale
($246$ GeV). I will use the name
{\it weak scale superstrings} to denote string
solutions with $m_s$ in the range from 250 GeV
up to a few TeV.

\newsec{An Example in Six Dimensions}

Weak scale superstrings are a subset of the class of
string solutions for which the ratio
$m_s/\mp$ can be tuned arbitrarily small while
keeping (at least some) gauge couplings of order one.
In six dimensions the gauge coupling has dimensions
of length; this defines an energy scale below
which the six-dimensional effective
gauge theory is weakly coupled. Thus one can
examine the six-dimensional analog of weak scale
superstrings by looking for solutions where
\eqn\wantsix{
{(\ap )^2\over \kappa ^2} \gg 1;\qquad
{\ap\over g^2} \sim {\rm O}(1)\quad .
}
where $\kappa$ is the six-dimensional gravitational
coupling.

There are two reasons for considering six-dimensional
examples first. One is that, given a six-dimensional
solution which satisfies \wantsix , we can in general
obtain four-dimensional solutions of the type we want
by further compactifying two dimensions
at a compactification scale which is of order one in string
units. More importantly, in six dimensions the constraints
from both anomaly cancellation and $N$$=$$1$ spacetime
supersymmetry are more severe than in four dimensions.
This allows one to extract information more reliably from
the interesting region of moduli space.

The first example I will discuss is a six-dimensional
compactification of the Type I superstring on a $K3$ $Z_2$
orbifold, a class of solutions recently constructed by
Gimon and Polchinski \polgim . These solutions have
$N$$=$$1$ (more precisely, (0,1)) spacetime supersymmetry,
the minimal amount of supersymmetry in six dimensions.
A toroidal compactification of such a solution to four
dimensions will produce solutions with $N$$=$$2$
supersymmetry.
In the case where all
sixteen of the Dirichlet 5-branes are at a fixed point of
the orbifold projection, the gauge group is
$U(16)$$\times$$U(16)$. The first/second $U(16)$
is carried by Chan-Paton factors associated with
open strings with ends attached to Dirichlet
9-branes/5-branes, respectively. Moving all sixteen
5-branes away from the fixed point and turning on
appropriate Wilson lines gives a very similar solution
with gauge group $USp(16)$$\times$$USp(16)$ \bsagnotti .

The massless particle content consists of the
gravity multiplet, one tensor multiplet, 20
gauge singlet hypermultiplets, the vector multiplets
of $U(16)$$\times$$U(16)$, and hypermultiplets
transforming under $U(16)$$\times$$U(16)$ as
a $(16,16)$, a $(120+\overline{120},1)$, and
a $(1,120+\overline{120})$.

Anomaly cancellation and spacetime supersymmetry fix
completely the form of certain terms in the effective
low energy field theory action
\refs{\green,\sagnotti,\dufw}. Thus in the
Einstein frame the action is:

\eqn\einact{
\eqalign{
{(2\pi )^3\over(\ap )^2} \int d^6x\, \sqrt{g}
\Big\{\; &R - {1\over 12}\etp H^2 \cr
&-{\ap\over 8}\sum _{\alpha =1,2} \left(
\va {\rm e}^{-\phi} + \vat {\rm e}^{\phi}
\right){\rm tr}\,F_{\alpha}^2 + \ldots\Big\} \quad .\cr
}}
Here $\phi$ is the scalar component of the tensor
multiplet, $R$ is the Ricci scalar, $H$ is the
3-form field strength, and $F_1$, $F_2$ are the
$U(16)$$\times$$U(16)$ field strengths.

Furthermore, the parameters $v_1$, $v_2$, ${\tilde v}_1$,
${\tilde v}_2$ are fixed by anomaly 

cancellation\footnote{$^*$}{For simplicity I will ignore 

the U(1) anomalies. For a complete analysis, see \newjoe .}.
The anomaly 8-form can be written \scharzsix :
\eqn\schanom{
I_8 = (\tr R^2)^2 + {1\over 6} \tr R^2 \sum_\alpha X_\alpha^{(2)} - {2\over 3}
\sum_\alpha X_\alpha^{(4)} + 4 \sum_{\alpha < \beta} Y_{\alpha\beta} ,
}
where
\eqn\moresch{
\eqalign{
X_\alpha^{(n)} &= \Tr F_\alpha^n - \sum_i n_i \tri F_\alpha^n
\cr
Y_{\alpha\beta} &= \sum_{ij} n_{ij} \, \tri F_\alpha^2 \, \trj F_\beta^2.
\cr
}}
Here the symbol $\Tr$ denotes a trace in the adjoint
representation and $\tri$ denotes a trace in the representation $R_i$ (of the
simple group $G_\alpha$).  $n_i$ is the number of hypermultiplets in the
representation $R_i$ of $G_\alpha$ and $n_{ij}$ is the number of
representations $(R_i, R_j)$ of $G_\alpha \times G_\beta$
which occur.
The Green-Schwarz anomaly cancellation mechanism requires
that the anomaly 8-form should factorize as
\eqn\factanom{
I_8 = (\tr R^2 - \sum_{\alpha} v_\alpha \tr F_\alpha^2)
(\tr R^2 - \sum_{\alpha} {\tilde v}_\alpha \tr F_\alpha^2)\quad ,
}
where $\tr$ denotes the trace in the fundamental representation.

Using the trace identities of ref. \erler , one finds
for the $U(16)$$\times$$U(16)$ model

\eqn\mysols{
\eqalign{
X_1^{(2)} &= -12 \tr F_1^2\quad ,\cr
X_2^{(2)} &= -12 \tr F_2^2\quad ,\cr
X_1^{(4)} &= X_2^{(4)} = 0\quad ,\cr
Y &= \tr F_1^2 \tr F_2^2 \quad .\cr
}}
Thus
\eqn\myfact{
I_8 = \left( \tr R^2 - 2\tr F_1^2 \right) \left(
\tr R^2 - 2\tr F_2^2 \right)
}
which implies:
\eqn\myparms{
\eqalign{
v_1 &=2 \; ,\qquad {\tilde v}_1 = 0 \cr
v_2 &=0 \; ,\qquad {\tilde v}_2 = 2 \cr
}}
The result $v_2$$=$$0$ indicates that the gauge
bosons of the second $U(16)$ are inherently
nonperturbative. This is expected as they are associated
with the Dirichlet 5-branes \refs{\dufw,\polnote}.

Let us now rescale from the Einstein frame to the
string metric frame; this is the frame in which
$m_s$ actually sets the scale of the Regge recurrences.
Rescale the metric by
\eqn\myresc{
g_{\mu\nu} \to {{\rm e}^{-\phi}\over \l} g_{\mu\nu}
}
where $\l$ is the ten-dimensional Type I string coupling.
Then \einact\ becomes:
\eqn\stringact{
\eqalign{
{(2\pi )^3\over(\ap )^4} \int d^6x\, \sqrt{g} \, V_I
\Big\{\; &{1\over \l ^2}R - {1\over 12}\etp H^2 \cr
&-{\ap\over 4\l}\tr F_1^2
-{(\ap )^3\over 4\l V_I}\tr F_2^2 + \ldots\Big\} \quad .\cr
}}
where
\eqn\whatv{
V_I \equiv \etp (\ap )^2
}
can be regarded as the effective compactification
volume; note this analysis in no way depends on an
implicit assumption that $V_I$ is large.

{}From \stringact\ we can read off the six-dimensional
gravitational and gauge couplings:
\eqn\hereyougo{
\eqalign{
{(\ap )^2\over\kappa ^2} &\sim {V_I \over \l ^2 (\ap )^2}\cr
{\ap\over g_1^2} &\sim {V_I\over \l (\ap )^2}\cr
{\ap\over g_2^2} &\sim {1\over \l}\cr
}}

The analog of weak scale superstrings thus
corresponds to very weak coupling and small $V_I$:
\eqn\myscaler{
\l \ll 1\; ,\qquad V_I/(\ap )^2 = {\rm O}(\l )
}
In this region of moduli space we then have:
\eqn\mystuff{
{(\ap )^2\over\kappa ^2} \sim {1\over\l} \; ,\quad
{\ap\over g_1^2} \sim {\rm O}(1) \; ,\quad
{\ap\over g_2^2} \sim {1\over\l} \; .
}
There are two widely separated energy scales.
The lower scale is the scale at which the first
Regge recurrences appear and at which the first
$U(16)$ gauge coupling gets strong. The higher
scale is the scale at which both gravity and the
second $U(16)$ gauge coupling get strong.
In this analogy the standard model gauge group
would be embedded in the first $U(16)$.

It is also instructive to look at the equivalent
heterotic or Type I$^{\prime}$  description of
these solutions. The table below show how the
string couplings and compactification scales are
related by duality \refs{\oldwit,\witpol}:

\eqn\firstrelate{
\eqalign{
{\rm Heterotic}\qquad\qquad
&{\rm Type\ I}\qquad\qquad\;{\rm Type\ I}^{\prime} \cr
\quad&\quad\cr
{1\over \lh}\qquad\qquad\quad\quad
&\;\l \qquad\qquad\;\;\;\quad {(\ap )^2\lp \over V_{I^{\prime}} }\cr
\quad&\quad\cr
\l ^2V_h\qquad\qquad\quad
&\; V_I \qquad\qquad\;\;\;\;\quad {(\ap )^4\over V_{I^{\prime}} }\cr
}}

In the heterotic description, we are in a region
of strong coupling and large radius. In the Type I$^{\prime}$
description we are also at large radius, but
the ten-dimensional string coupling is
of order one.

\newsec{Another Example in Six Dimensions}

Another simple example comes from the $SO(32)$ heterotic
string compactified on $K3$. The $K3$ compactification
requires a gauge bundle with instanton number 24.
As shown by Witten \witan , at the special region in
moduli space where all 24 instantons
shrink to zero size, the gauge group is enhanced
to $SO(32)$$\times$$Sp(24)$. The extra $Sp(24)$ gauge
bosons are inherently nonperturbative and are associated
with solitonic 5-branes, just as the second $U(16)$ in
the $K3$ orbifold discussed above was associated with
the dual Dirichlet 5-branes.

Because of anomaly cancellation and supersymmetry
the low energy effective action in the Einstein
frame has the same form as \einact . The
$v$, $\tilde v$ parameters are determined to
be \swsix :
\eqn\swparms{
\eqalign{
v_{32} &=1 \; ,\qquad {\tilde v}_{32} = -2 \cr
v_{24} &=0 \; ,\qquad {\tilde v}_{24} = 2 \cr
}}

For the heterotic string e$^{\phi}$ is the
six-dimensional effective string coupling, i.e.
\eqn\whatiz{
{\rm e}^{2\phi} = {\lh ^2\over V_h}
}
where $V_h$ is the volume of $K3$ and $\lh$ is
the ten-dimensional string coupling.
Thus the proper rescaling from the Einstein frame to the
string metric frame is given by:
\eqn\myresch{
g_{\mu\nu} \to {\rm e}^{-2\phi} g_{\mu\nu}
}
Then \einact\ becomes:
\eqn\hstringact{
\eqalign{
{(2\pi )^3\over(\ap )^4} \int d^6x\, \sqrt{g} \, V_h
\Big\{\; &{1\over \lh ^2}R - {1\over 12\lh ^2} H^2 \cr
&-{\ap\over 8\lh ^2}\left( 1 - {2(\ap )^2\lh ^2\over V_h}
\right)\tr F_{32}^2
-{(\ap )^3\over 4 V_h}\tr F_{24}^2 + \ldots\Big\} \quad .\cr
}}

{}From \hstringact\ we can read off the six-dimensional
gravitational and gauge couplings:
\eqn\hereyougoh{
\eqalign{
{(\ap )^2\over\kappa ^2} &\sim {V_h \over \lh ^2 (\ap )^2}\cr
{\ap\over g_{32}^2} &\sim {V_h \over \lh ^2 (\ap )^2} - 2 \cr
{\ap\over g_{24}^2} &\sim 1\cr
}}

Let us then consider the case where
the ten-dimensional string coupling $\lh $ is of order
one, while the heterotic volume $V_h$ is large.
In this region of moduli space we then have:
\eqn\mystuff{
{(\ap )^2\over\kappa ^2} \sim {V_h\over (\ap )^2} \; ,\quad
{\ap\over g_{32}^2} \sim {V_h\over (\ap )^2} \; ,\quad
{\ap\over g_{24}^2} \sim 1 \; .
}
There are two widely separated energy scales.
The lower scale is the scale at which the first
Regge recurrences appear and at which the
$Sp(24)$ gauge coupling gets strong. The higher
scale is the scale at which both gravity and the
$SO(32)$ gauge coupling get strong.
In this analogy the standard model gauge group
would be embedded in $Sp(24)$.

\newsec{Realistic Weak Scale Superstrings}

The six-dimensional examples considered above
are very far from a solution which could
correspond to a realistic weak scale superstring.
One obvious difficulty is that taking the compactification
volume to be very large in string units (as in the
second example or in the Type I$^{\prime}$ picture
of the first example) is a phenomenological disaster if
the string scale itself is only a TeV. Thus for a
realistic model we must suppose that the small ratio
$m_s/\mp$ is associated with some modulus which can
get a very large vev without generating unwanted observable light states.

Another obvious difficulty is the notorious problem
of stablizing the vev of the dilaton \oldsd .
In any weak coupling limit of the superstring,
the dilaton vev vanishes -- another phenomenological
disaster. As discussed by Dine and Shirman \dinshir ,
a realistic superstring probably must reside in a region
of moduli space which admits no weak coupling
description.
Both six-dimensional examples
fail this criterion, the first in the Type I description
and the second in the Type I$^{\prime}$ description.
However this failure is not as bad as it could
have been, since in both cases
the weak coupling, small radius
description is only accessible due to extra
symmetry of the compactifications.
In a realistic solution we should at any rate
avoid extra symmetries which can prevent
a stable nonzero dilaton vev even at intermediate
and strong coupling \bandin .

Dine and Shirman \dinshir\ have identified a
possibly unique
region of the moduli space of four-dimensional compactifications
which satisfies their criterion without
requiring all moduli to take intermediate values.
This ``truly strong coupling'' region corresponds
to $\lh \gg 1$, with the 6-dimensional compactification
volume $V_h$ scaling like $\lh ^2$. Thus in the heterotic,
Type I, and Type I$^{\prime}$ pictures we have:

\eqn\relatemore{
\eqalign{
{\rm Heterotic}~~~~~\qquad\qquad
&{\rm Type\ I}\qquad\qquad\;\quad\qquad{\rm Type\ I}^{\prime} \cr
\quad&\quad\cr
\lh\qquad\qquad\quad\quad
&\;\l = {1\over\lh}\qquad\qquad
\;\;\;\qquad\lp = {\rm O}(1) \cr
\quad&\quad\cr
V_h \sim \lh ^2\qquad\qquad\quad
&\; V_I = {\rm O}(\l ) \qquad\qquad\;\;
V_{I^{\prime}}={\rm O}(1/\l ) \cr
}}

Consider such solutions in the Type I description. If
the 6-volume $V_I$ were large instead of small, we
would be justified in writing the effective action as:
\eqn\act{
{(2\pi )^3\over(\ap )^4} \int d^4x\, \sqrt{g} \, V_I
\Big\{\; {1\over \l ^2}R
-{k\ap\over 4\l}\tr F^2 + \ldots\Big\} \quad .
}
where $k$$=$$1$ for large volume.
For weak coupling and large volume, we can
read off the gauge and gravitational
couplings from the tree-level terms in \act .
As $V_I$ shrinks, this is
no longer true, in general. In fact for
$V_I$$\sim$$\l$$\ll$$1$, one should regard
$V_I$ as representing the scaling of some moduli,
but not as a classical volume.

However there is
likely to be a large subclass of solutions
in the ``truly strong coupling'' region
where the gauge and gravitational couplings
are still determined by an effective action
of the form \act , where $V_I$ is to be regarded
as some scaling function of moduli and the parameter
$k$ (also a function of some moduli) is of order one.
As discussed above, we also must require that the modulus
vev that makes $V_I$ small must somehow not
also lead to unwanted observable light states. Whether
this is likely --or even possible-- I do not know.

For these solutions we will have $\ag$ of order one
while
\eqn\mytrelev{
\ap \mp ^2 = {4\over k\l\ag }
}
i.e. the string scale is arbitrarily smaller than the
Planck scale. Thus the ``truly strong coupling'' region
(broadly interpreted) may be a likely place to find
realistic weak scale superstrings, if they exist.

\newsec{Objections to Weak Scale Superstrings}

\subsec{Gauge coupling unification}

In reference \newwit\ the results summarized
in the introduction were obtained
in the context of obtaining a modest reduction in
the ratio $m_s^2/\mp ^2$ beyond what is implied by
\trelev . The motivation is the well known apparent
gauge coupling unification at $\sim 10^{16}$ GeV
implied by a naive renormalization group evolution
of the measured low energy couplings plus minimal
SUSY thresholds.

However it is not at all obvious that gauge coupling
unification in the string sense has any direct
relation to this apparent unification of the standard
model gauge couplings. Even for the heterotic string
at weak coupling, we know (see the discussion
below \trelev ) that gauge coupling unification in
the string sense does not necessarily imply
{\it equality} of the gauge couplings at some scale.
Thus the only argument pinning the string scale
to $10^{16}$ GeV is the conviction that the
apparent unification at that scale is ``too close''
to be a coincidence. This argument is even weaker
than it seems, since it is possible that, while
{\it not} a coincidence, the apparent unification maps
into some sophisticated structure of the underlying
string theory, without requiring an {\it actual} field
theory desert between $10^3$ and $10^{16}$ GeV.

\subsec{The success of weak coupling heterotic models}

A number of weak coupling heterotic string models have
been built which exhibit an elegant confluence of
favorable phenomenological attributes. These models
have three generations of standard model chiral fermions,
embed the standard model gauge group, and have a natural
hidden sector suitable for dynamical supersymmetry breaking.
They also exhibit new symmetries which naturally give
a hierarchical structure to the Yukawa matrices.
For recent reviews, see \refs{\ferr,\keithr,\jorr,\farman,\mer}.

Thus one could argue that the hypothesis of
weak scale superstrings moves us very far away
from a class of string solutions which look very
much like the real world.

One problem with this argument is that it
includes a number of theoretical assumptions in its
definition of ``the real world''. Another problem is that 

we are only just beginning to understand the principles
which control the relationships between phenomenological attributes 

in such solutions. Some features of these solutions,
such as symmetries of the superpotential which
restrict Yukawa couplings, should survive if
we deform the solutions into the intermediate coupling
region \bandin , where (we hope) the dilaton vev
is stabilized. But beyond this it is still
premature to use these weak coupling solutions
as a way of constraining properties of a realistic string solution.

\subsec{Spacetime supersymmetry}

Spacetime supersymmetry is motivated in
particle theory as a way to stabilize the
hierarchy between the electroweak scale
and the Planck scale. With superstrings, this ties in nicely with the
fact that spacetime SUSY also removes tachyons
from the physical string spectrum, and
guarantees a vanishing cosmological constant.

If the string scale is around a TeV we lose the
original motivation for spacetime SUSY. In
fact spacetime supersymmetry becomes a serious
problem, since it is notoriously
difficult to break supersymmetry in a
phenomenologically acceptable way at such
a low scale. Furthermore the supersymmetry
mass splittings would now be the same order
of magnitude as the spacing of the Regge
recurrences.

This suggests that a viable weak scale
superstring solution may {\it not} exhibit
spacetime supersymmetry
in the effective field theory below the string scale.

\subsec{Why the electroweak scale?}

Why the Planck scale? String dynamics softens the
ultraviolet behavior of quantum gravity. With
the possible exception of cosmology, I know of no
consideration which says that these stringy effects
cannot set in at a scale where gravitational forces
are still weak. Of course, since the low energy
effective field theory action will contain
an infinite number of higher dimension terms
suppressed by powers of the string scale, weak
scale superstrings are constrained somewhat by
low energy data -e.g. flavor changing neutral
currents. But these constraints are no more
severe than for other new physics scenarios at
the TeV scale.

\newsec{Experimental Consequences}

The hypothesis of weak scale superstrings has spectacular
consequences for collider physics at TeV energies.
Each of the known particles of the standard model
(as well as the graviton) sits at the base of a
Regge trajectory. There are an infinite number of
Regge recurrences, with progressively higher masses and
spins. These particles carry standard model
quantum numbers including color and are unstable. The
lightest ones could have masses as low as a few hundred
GeV without violating current experimental bounds.

An obvious guess for the lightest Regge recurrences
are the heavy spin 3/2 partners of light quarks and leptons.
For masses in the range from a few hundred GeV to a TeV
the heavy spin 3/2 quarks will be easier to detect than
the heavy leptons.

The relatively light Regge recurrences may also be
accompanied by relatively light Kaluza-Klein modes,
if one or more of the effective compactification
radii is of order the weak scale rather than the
Planck scale. In this case \ankk\ a plausible guess for
the lightest Kaluza-Klein modes are the heavy partners
of the gluons.

In this regard it is interesting to note that
either spin 3/2 heavy quarks or heavy color octets
are possible explanations \refs{\band,\bagking}
of the excess in jet production
for ${\rm E}_t$$>$ $200$ GeV
reported by the CDF collaboration in $p\bar p$
collisions at the Tevatron \cdfjet .

The effects of Regge recurrences on
the single jet inclusive cross section will
resemble the effects of compositeness: in both
cases the amplitude has an $s/M$ enhancement
at high ${\rm E}_t$. However it should be possible
to distinguish the higher spin Regge recurrences
by examining the jet angular distributions.

If the real world is a weak scale superstring
the LHC will produce unintelligible results when
operated at design energy and luminosity.
It will be necessary in that case to resort
to something like a DiTevatron or TEV33 to have any hope
of sorting out the superstring
threshold region.

Weak scale superstrings also have profound
implications for cosmology and black hole physics.
The number of
heavy string states increases exponentially with
mass; this implies a Hagedorn
temperature of a few TeV \hag .
The existence of such a Hagedorn transition will require
a radical rethinking of inflation, structure
formation, and baryogenesis.

It will be difficult to construct
realistic weak scale superstring models,
even if they exist. But if they are there,
we will certainly discover them in high energy colliders.

\listrefs

\bye